\documentclass[10pt,letterpaper,twocolumn]{article} 

\usepackage{ol2}
\usepackage[draft]{hyperref}
\usepackage{amsmath}

\begin{document}

\twocolumn[ 

\title{Non-Hermitian topological mobility edges and transport in photonic quantum walks}


\author{Stefano Longhi}
\address{Dipartimento di Fisica, Politecnico di Milano and Istituto di Fotonica e Nanotecnologie del Consiglio Nazionale delle Ricerche, Piazza L. da Vinci 32, I-20133 Milano, Italy (stefano.longhi@polimi.it)}
\address{IFISC (UIB-CSIC), Instituto de Fisica Interdisciplinar y Sistemas Complejos, E-07122 Palma de Mallorca, Spain}

\begin{abstract}
 In non-Hermitian quasicrystals, mobility edges (ME) separating localized and extended states in complex energy plane can arise as a result of non-Hermitian terms in the Hamiltonian. Such ME are of topological nature, i.e. the energies of localized
and extended states exhibit distinct topological structures in the complex energy plane. However, depending on the origin of non-Hermiticity, i.e. asymmetry of hopping amplitudes or complexification of the 
incommensurate potential phase, different winding numbers are introduced, corresponding to different transport features in the bulk of the lattice: while ballistic transport is allowed in the former case,  pseudo dynamical localization is observed in the latter case. The results are illustrated by considering non-Hermitian photonic quantum walks in synthetic mesh lattices. 
\end{abstract}

 ] 

{\it Introduction.} Mobility edges (ME), separating localized from extended states, are known to arise in the single-particle energy spectrum of three-dimensional disordered lattices (like in the Anderson model), and in certain one-dimensional lattices with an incommensurate potential (quasicrystals) \cite{M1,M2,M3}. In Hermitian systems, a rather general (though not universal \cite{M4}) rule is that the existence of extended states and absolutely continuous spectrum implies  dynamical delocalization and ballistic transport. Hence in case of a ME dynamical delocalization is commonplace. 
 Recently, topological phases in non-Hermitian (NH) systems have sparked a great interest (see e.g. \cite{M5,M6,M7,M8,M9,M10,M11,M12,M13,M14,M15,M16,M17,M18,M19,M20,M21,M21b,M22} and references therein), with the prediction and observation of exotic phenomena, such as enriched topological
classification \cite{M5,M11}, the breakdown of the conventional bulk-boundary correspondence \cite{M6,M7,M9,M13,M14,M19} and the NH skin
effect \cite{M7,M10,M14,M15,M19}. In particular, exciting physics arises from the interplay among topology, non-Hermiticity and aperiodic order \cite{M23,M24,M25,M26,M27,M28,M29,M30,M31,M32,M33,M34,M35,M36,M37,M38,M39,M40}. Interestingly, in NH quasicrystals the metal-insulator phase transition and the appearance of ME induced by the non-Hermiticity are of topological nature, as they can be traced back to the change of a spectral winding number \cite{M5,M23,M29,M31,M33,M35,M36}. From the experimental side, photonic quantum walks in synthetic lattices  have provided a fantastic platform for the experimental demonstration of such exotic effects \cite{M13,M16,M21,M39,M40}.\par
 In this Letter we study the dynamical (transport) properties in quasicrystals where topological ME are driven by non-Hermiticity, and show that, unlike Hermitian systems, ballistic transport can be prevented in the NH case. In particular, two different types of topological ME can be created in NH quasicrystals: by either assuming asymmetric hopping amplitudes (type-I ME) or by complexification of the 
incommensurate potential phase (type-II ME). While ballistic transport is allowed in type-I ME,  it is forbidden in type-II ME, where a regime of pseudo localization is observed. These results are illustrated by considering discrete-time photonic quantum walks in synthetic mesh lattices with a bichromatic quasi-periodic potential.\\
{\it Non-Hermitian mobility edges and transport in quasicrystals.}  We consider a one-dimensional NH quasicrystal with nearest-neighbor hopping described by the single-particle Hamiltonian $\hat{H}=\sum_{n,m} H_{n,m} |n \rangle \langle m|$ with matrix Hamiltonian $H$ given by
\begin{equation}
H_{n,m}=\kappa \exp(h+i \theta) \delta_{n,m-1}+\kappa \exp(-h-i \theta) \delta_{n,m+1}+V_n \delta_{n,m}
\end{equation}
where $V_n$ is the incommensurate potential, $\kappa$ the hopping amplitude, and $h+i \theta$ a complex Peierls (gauge) phase. The on-site potential $V_n$ is given by $V_n=V(x=2 \pi \alpha n+ \varphi+i \epsilon)$, where $V=V(x)$ is a $2 \pi$-periodic real function, $\alpha$ an irrational Diophantine number and $\varphi+i \epsilon$ the complex potential phase. To highlight the dependence of the Hamiltonian $H$ on either the phases $\theta$ or $\varphi$, we will write $H=H(\nu)$ with $\nu=\theta, \varphi$. We assume a finite lattice comprising a sufficiently large number $L$ of sites in a ring geometry, so that periodic boundary conditions (PBC) apply. The lattice size $L$ is taken so as to provide a rational approximant of the irrational $\alpha$. For convenience, we assume $\alpha=(\sqrt{5}-1)/2=\lim_{n \rightarrow \infty} (F_{n-1}/F_{n})$ (the inverse of the golden mean) and $L=F_n$ for large enough $n$, where $F_n$ are the Fibonacci numbers defined recursively by $F_{n+1} = F_n + F_{n-1}$ and $F_0 = F_1 = 1$.
The matrix Hamiltonian $H$ is Hermitian for $\epsilon=h=0$. Non-Hermiticity can be thus introduced in two ways:  (i) by allowing for a non-vanishing imaginary gauge field $h \neq 0$, corresponding to asymmetric hopping amplitudes \cite{M41}; and (ii) by considering a complex potential phase $\epsilon \neq 0$ \cite{M23}. Let us now suppose that in the Hermitian limit $h=\epsilon=0$ the Hamiltonian does not possess ME, i.e. all eigenstates are either exponentially-localized or extended. In the former case, ME can be induced rather generally by the introduction of an imaginary gauge field ($h \neq 0$) via the mechanism dubbed non-Hermitian delocalization transition \cite{M5,M41}: some localized eigenstates become delocalized and acquire complex energies, while the other eigenstates remain localized with real and unchanged energies. Such a kind of NH-induced ME is referred to as type-I ME. We note that such systems display rather generally the NH skin effect, so as the energy spectrum is strongly sensitive to the boundary conditions and ME disappear under OBC: therefore, we will consider here PBC solely. When all eigenstates of $H$ in the Hermitian limit are extended, ME can be induced by a complex phase $\epsilon \neq 0$ in the incommensurate potential, so as some extended states become localized with complex energies, while the other eigenstates remain extended with real and unchanged energies. Such a kind of NH-induced ME is referred to as type-II ME. The physical mechanism underlying the formation of type-II ME is again the non-Hermitian delocalization transition but for the lattice in the reciprocal (Fourier) space, where the complex phase $\epsilon$ in the potential acts as an imaginary gauge field \cite{M23} . The formation of NH type-I and type-II ME is illustrated in Fig.1 for the quasi-periodic potential $V(x)= \lambda_1 \cos(x)+ \lambda_2 \cos(2x)$, corresponding to a generalized Aubry-Andr\'e model. This model is feasible for an experimental implementation and ensures the appearance of NH ME for $\lambda_{1,2} \neq 0$, owing the breakdown of the self-duality of the Aubry-Andr\'e model.
Localized and extended wave functions $\psi_n$ are identified by their inverse participation ratio  $IPR= \sum_n | \psi_n|^4 /( \sum_n| \psi_n|^2 )^2$, with $IPR \sim 1$ for a tight localized state and $IPR \sim 1/L \ll 1$ for an extended state. In Figs.1(a) and (b), the potential amplitudes $V_1$ and $V_2$ are set so as in the Hermitian limit all wave functions are localized. Application of an imaginary gauge field $h$ clearly creates a ME with extended wave functions corresponding to complex energies forming closed loops in complex energy plane, while localized wave functions retain their real energies like in the Hermitian case. In Figs.1(c) and (d) the potential amplitudes $V_1$ and $V_2$ are set so as in the Hermitian limit all wave functions are extended. Application of an imaginary phase $\epsilon$ clearly creates a ME with localized wave functions corresponding to complex energies forming closed loops in complex energy plane while extended wave functions retain their real energies. 
As shown in previous works \cite{M5,M23,M29,M33,M34,M35,M36}, both types of ME are of topological nature, in the sense that the energies of localized
and extended states exhibit distinct topological structures in the complex energy plane. Specifically, for a given base energy $E_B$ the following two spectral winding numbers can be introduced
\begin{equation}
W_{\nu}= \lim_{L \rightarrow \infty} \frac{1}{2 \pi i L} \int_0^{2 \pi} d \nu \frac{d}{d \nu} \log \left\{ \det (H(\nu)-E_B)  \right\}
\end{equation}
  \begin{figure}[htb]
\centerline{\includegraphics[width=8.5cm]{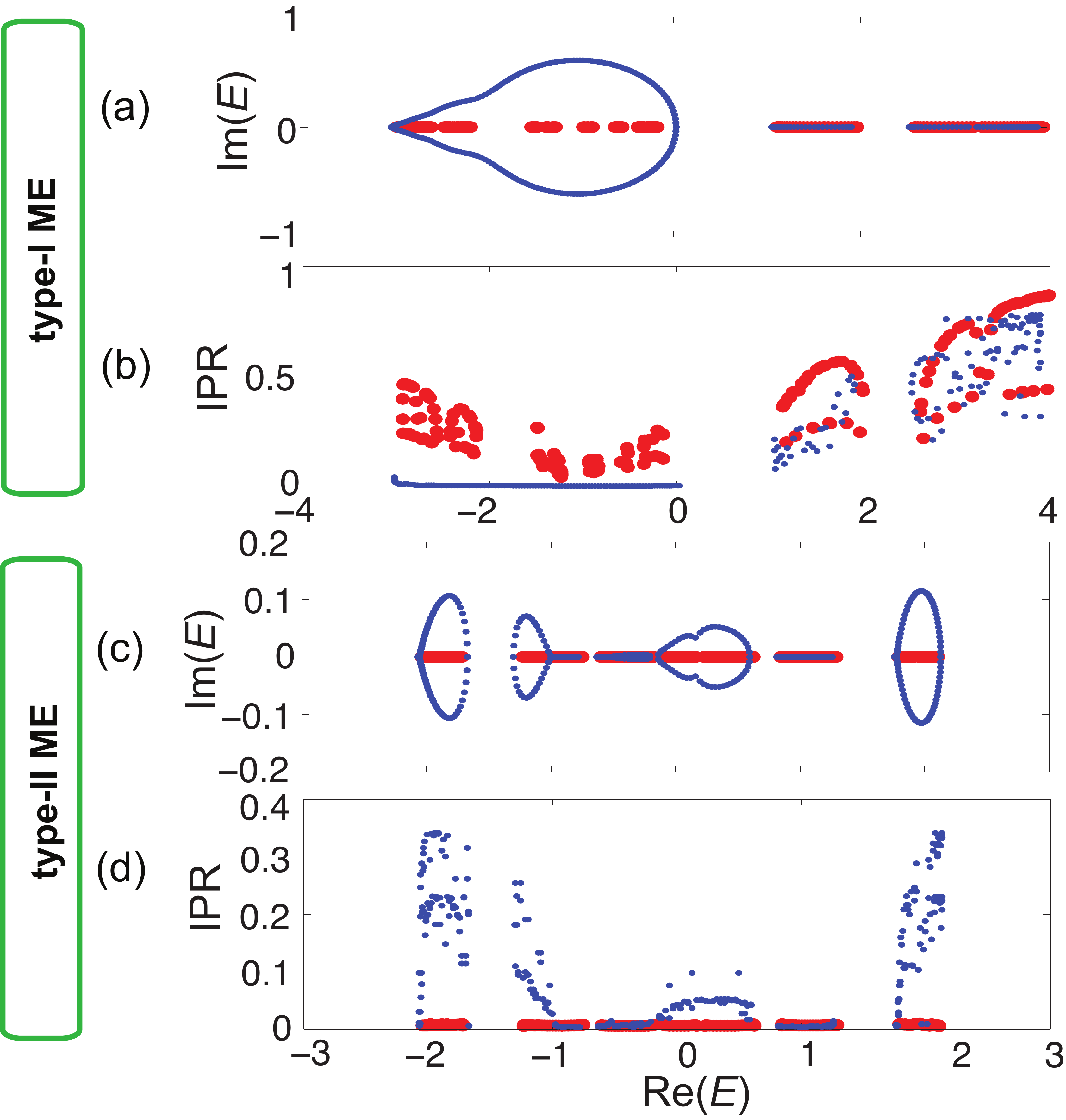}} \caption{ \small
(Color online) (a,c) Energy spectrum and (b,d) IPR of eigenfunctions of a quasicrystal in a ring geometry with the incommensurate bichromatic potential $V(x)=V_1 \cos(x)+V_2 \cos(2x)$. In (a,b), $V_1=2$ and $V_2=1.5$, corresponding to all states localized in the Hermitian limit ($\epsilon=h=0$; red bold dots). Application of the imaginary gauge field ($h=0.5$, $\epsilon=0$; blue dots) creates a type-I ME: some states become delocalized, with energies describing a closed loop in complex plane, while the other states remain localized and retain their original real energies. In (c,d), $V_1=0.2$ and $V_2=0.5$, corresponding to all states extended in the Hermitian limit ($\epsilon=h=0$; red bold dots). Application of the imaginary potential phase ($\epsilon=0.6$, $h=0$; blue dots) creates a type-II ME: some states become localized, with energies describing closed loops in complex plane, while the other states remain extended and retain their original real energies. Other parameter values are $\kappa=1$, $\alpha=(\sqrt{5}-1)/2$, $\theta=\varphi=0$ and $L=377$.}
\end{figure} 
where $\nu$ is either $\theta$ or $\varphi$. In type-I ME, the relevant winding number is $W_{\theta}$, which takes a non-vanishing value whenever $E_B$ is internal to any one of the complex energy loops associated to the extended eigenstates, while $W_{\varphi}$ is always vanishing. Likewise, in type-II ME the relevant winding number is $W_{\varphi}$, which is non-vanishing whenever $E_B$ is internal to any one of the complex energy loops associated to the localized eigenstates while $W_{\theta}$ is always vanishing.\\
While the existence of different types of ME has been previously predicted \cite{M32}, it is a full open question whether the different nature of the two types of ME, characterized by the two distinct topological numbers, can be unveiled by considering the {\em dynamical localization} properties of the system, i.e. the possibility of an initially localized excitation to spread or move in the lattice. Different types of transport in NH systems, displaying or not the NHSE, have been observed in previous works \cite{M13,M42}, however such previous systems did not show any ME. To disclose the interplay between transport features and NH ME,
let us indicate by $\psi_n(t)$ the occupation amplitude of the lattice at site $n$ and time $t$ for the initial condition, at time $t=0$, corresponding to the excitation of the single site at $n=n_0$, i.e. $\psi_n(0)= \delta_{n,n_0}$, and let $\tilde \psi_n(t)=\psi_n(t) / \sqrt{\sum_n | \psi_n(t)|^2}$ be the corresponding normalized occupation amplitudes. As a measure to characterize wave spreading in the lattice, we consider the second-order moment \cite{M37,M42}
\begin{equation}
\sigma^2(t)=\sum_n (n-n_0)^2 | \tilde {\psi}_n(t)|^2
\end{equation}
for the position operator. Ballistic transport corresponds to an asymptotic linear growth in time $\sigma(t) \sim v t$ with some characteristic speed $v$; diffusive transport corresponds to a lower-than-linear growth $\sigma(t) \sim t^{\delta}$ ($ \delta <1$); and dynamical localization corresponds to a bounded behavior of $\sigma(t)$ with time. While there is not a one-to-one correspondence between spectral
and dynamical properties of a quantum system, a general rule of thumb in an Hermitian system is that the absolutely continuous spectrum, corresponding to extended states, yields ballistic transport at a speed $v$ 
which is proportional to the Lebesgue measure of the absolutely continuous spectrum \cite{M37}.  However, in NH systems such a scenario is deeply changed since the energies of localized and/or extended states can acquire a complex energy, thus introducing different lifetimes of various wave functions \cite{M42}. Figure 2 shows typical numerical results of wave spreading in the NH quasicrystals with either type-I [Fig.2(a)] or type-II [Fig.2(b)] ME. In the former case, $\sigma(t)$ grows linearly with time, indicating ballistic transport like in the Hermitian limit. The transport is allowed by the extended states, which acquire complex energies and dominate the dynamics at long times. As compared to the Hermitian case, transport is unidirectional owing to the convective motion introduced by the imaginary gauge field \cite{M10}. A very different scenario is observed in type-II ME, where localized states acquire complex energies while extended states have real energies. In this case the spreading in the lattice is greatly suppressed and a pseudo-localization regime is observed [Fig.2(b)]. Interestingly, in the early stage of the dynamics (up to $t \sim 30$) one observes some fast wave spreading which is the signature of extended states with real energies. Subsequently, the spreading is stopped and dynamical localization is observed for a while. On a much longer time scale, a further increase of $\sigma(t)$ can be eventually observed, corresponding to a jump of the excitation to a different lattice region. Such a behavior stems form the fact that in type-II ME localized eigenstates of $H$ have a larger growth rate (imaginary part of the energy) than extended states. Hence, at longer time they dominate the dynamics, leading to a regime of pseudo-localization with stochastic jumps among localized eigenstates possessing different lifetimes, a regime similar to the one recently observed in Ref.\cite{M42}.\\
  \begin{figure}[htb]
\centerline{\includegraphics[width=8.7cm]{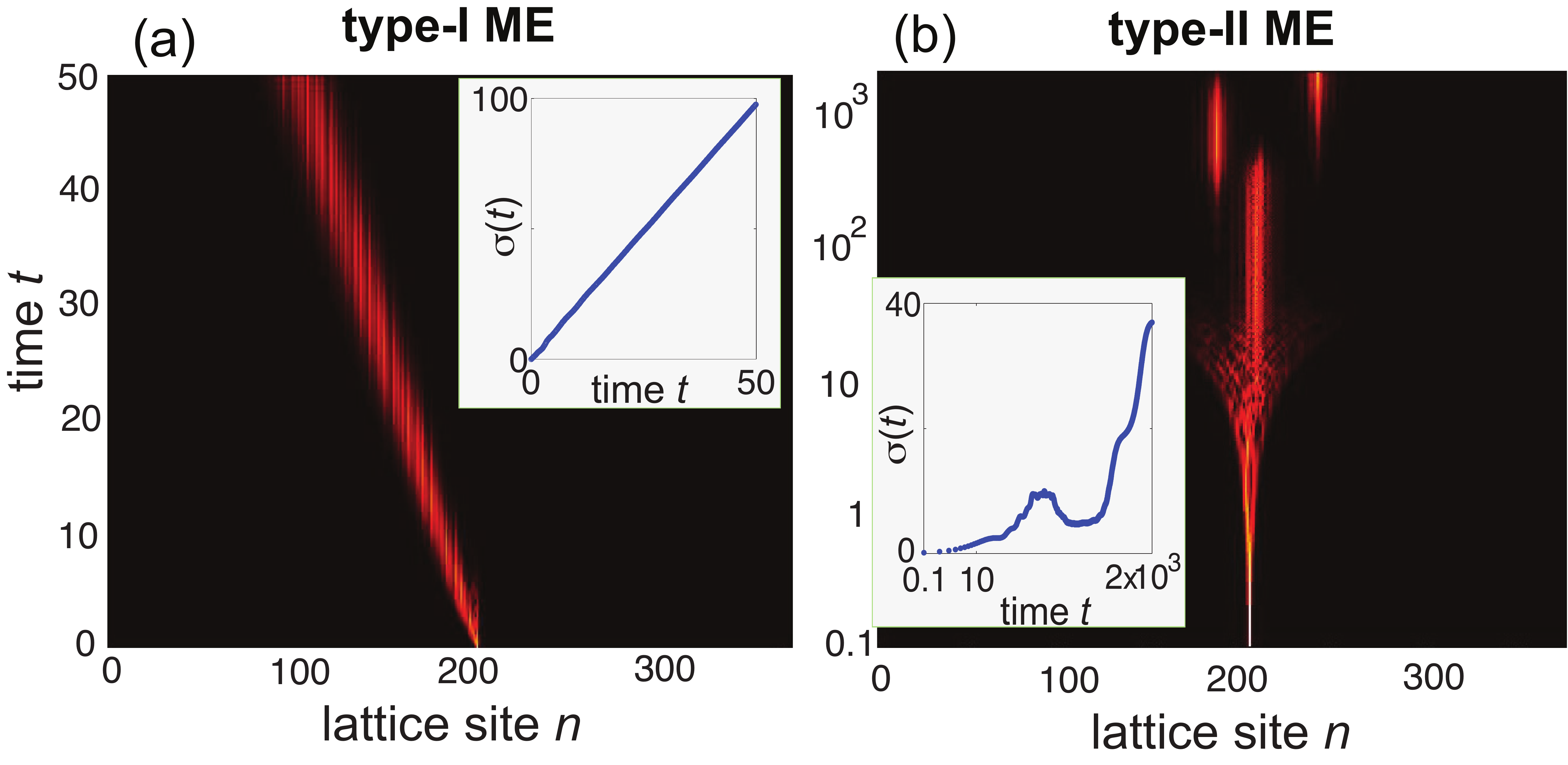}} \caption{ \small
(Color online) Wave spreading in a NH quasicrystal with (a) type-I, and (b) type-II ME. Parameter values in the two cases are the same as in Fig.1. The panels show, on a pseudo color map, the temporal evolution of the normalized occupation amplitudes $|\tilde{\psi}_n(t)|$ for single-site excitation of the lattice at $t=0$. The insets depict the corresponding behavior of the second-order moment $\sigma(t)$. Note in (b) the use of a log scale for time.}
\end{figure} 
 {\em Mobility edges in photonic quantum walks.} The main results discussed above for the continuous time evolution of the wave function governed by the NH Hamiltonian $H$ can be extended to discrete-time quantum walks, which have provided a feasible platform to experimentally observe the exotic properties of NH synthetic crystals and quasicrystals \cite{M13,M16,M21,M39,M40,M42}. Specifically, let us consider a photonic quantum walk realized in coupled optical fiber loops \cite{M16,M39,M42,M43}, where the imaginary gauge field $h$ is introduced by balanced gain/loss  in the two fiber loops while a complex incommensurate potential $V(x)$ is emulated by using synchronized amplitude and phase modulators in one of the two fiber loops. Light dynamics  is governed by the discrete-time coupled equations \cite{M16,M39,M42,M43}
 \begin{eqnarray}
 u^{(m+1)}_n & = & \left[   \cos \beta u^{(m)}_{n+1}+i \sin \beta v^{(m)}_{n+1}  \right]  \exp (h-2iV_n) \\
 v^{(m+1)}_n & = & \left[   \cos \beta v^{(m)}_{n-1}+i \sin \beta u^{(m)}_{n-1}  \right] \exp (-h)
 \end{eqnarray}
 where $u^{(m)}_n$, $v_n^{(m)}$ are the pulse amplitudes at lattice position $n$ and at discrete time step $m$ on the left and right moving paths, respectively, $\beta$ is the coupling angle of the beam splitter, and $V_n=V(x=2 \pi \alpha n+\varphi+i \epsilon)$ is the incommensurate potential. A connection between the discrete-time quantum walk dynamics [Eqs.(4) and (5)] and the continuous-time dynamics  described by the Hamiltonian $H$ [Eq.(1)] can be established in the limit $\beta \simeq \pi/2$ and $|V_n| \ll 1$, as shown in the Supplemental document. In this limit, the discrete-time quantum walk dynamics basically splits into two independent  continuous-time processes described by two Hamiltonians $H_{\pm}$ given by Eq.(1) with an incommensurate potential $V_n$ and with opposite effective hopping amplitude $ \kappa= \pm (\pi/2 - \beta)/2$. The appearance of the two types of ME and corresponding transport properties discussed in previous section are therefore expected to be observable in discrete-time NH photonic quantum walks. As an example, Fig.3 shows the numerically-computed quasi energy spectra (under PBC) and IPR of corresponding eigenstates in discrete-time quantum walks for the bichromatic incommensurate potential $V(x)=V_1 \cos(x)+V_2 \cos(2x)$ and for parameter values corresponding to the appearance of type-I [Fig.3(a,b)] and type-II [Fig.3(c,d)] ME. Note that the quasi energy spectra are grouped into two blocks, spaced by $\pi$ and mapping the eigenenergies of the two Hamiltonians $H_{_{\pm}}$ in the continuous-time limit (see Supplemental document).
     \begin{figure}[htb]
\centerline{\includegraphics[width=8.7cm]{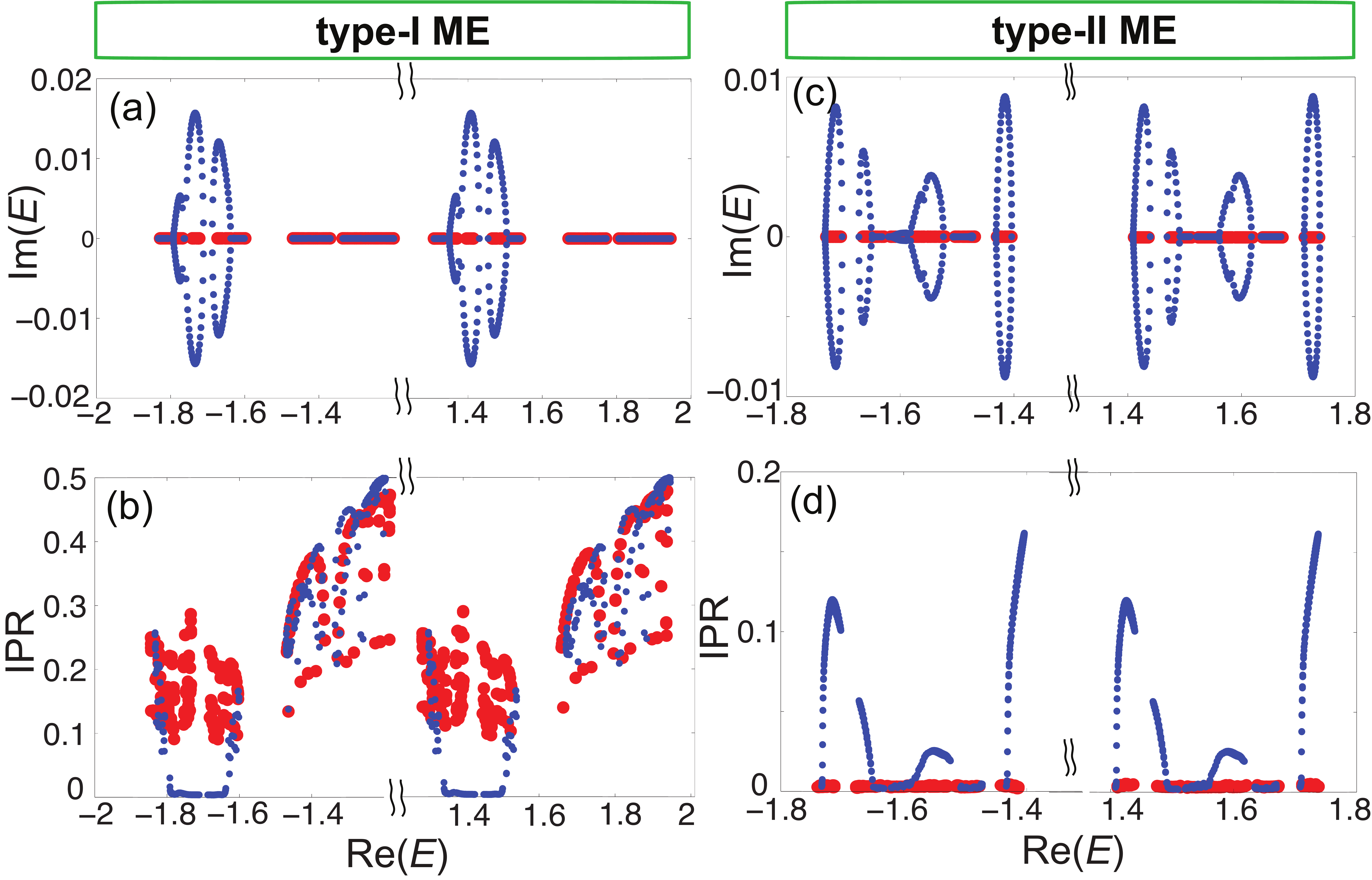}} \caption{ \small
(Color online) (a,c) Quasi energy spectrum and (b,d) IPR of eigenfunctions in the NH discrete-time photonic quantum walk for the bichromatic incommensurate potential  $V(x)=V_1 \cos(x)+V_2 \cos(2x)$ and for $\beta= 0.9 \times (\pi /2)$. In (a,b), $V_1=0.2356$ and $V_2=0.1178$, corresponding to all localized states in the Hermitian limit ($\epsilon=h=0$; red bold dots). Application of the imaginary gauge field ($h=0.4$, $\epsilon=0$; blue dots) creates a type-I ME. In (c,d), $V_1=0.0157$ and $V_2=0.0393$, corresponding to all extended states in the Hermitian limit ($\epsilon=h=0$; red bold dots). Application of the imaginary potential phase ($\epsilon=0.6$, $h=0$; blue dots) creates a type-II ME. Note that the quasi energy spectra are grouped into two blocks spaced by $\pi$, as discussed in the main text. Other parameter values are $\alpha=(\sqrt{5}-1)/2$, $\varphi=0$, and $L=377$.}
\end{figure} 
  The different dynamical properties of the two types of MEs are shown in Fig.4. Like in the continuous-time Schr\"odinger equation, we excite the lattice in a single site by launching a single optical pulse in one of the two fiber loops, namely we numerically solve Eqs.(4) and (5) with the initial condition $u_n^{(0)}= \delta_{n,n_0}$ and $v_{n}^{(0)}=0$. The spreading of the excitation in the synthetic lattice versus discrete time $m$ is measured by the second-order moment $\sigma^2(m)= \sum_n (n-n_0)^2 | \tilde{\psi}_n^{(m)} |^2 $, where we have set $| \tilde{\psi}_n^{(m)}|^2= \left( |u_n^{(m)}|^2+ (|v_n^{(m)}|^2 \right)  / \sum_n \left( |u_n^{(m)}|^2+ |v_n^{(m)}|^2  \right) $. Note that for type-I ME unidirectional ballistic transport is observed [Fig.4(a)], while a pseudo-localization regime, like the one discussed in Fig.2(b), is observed for type-II ME [Fig.4(b)].\\
 \par

  \begin{figure}[htb]
\centerline{\includegraphics[width=8.7cm]{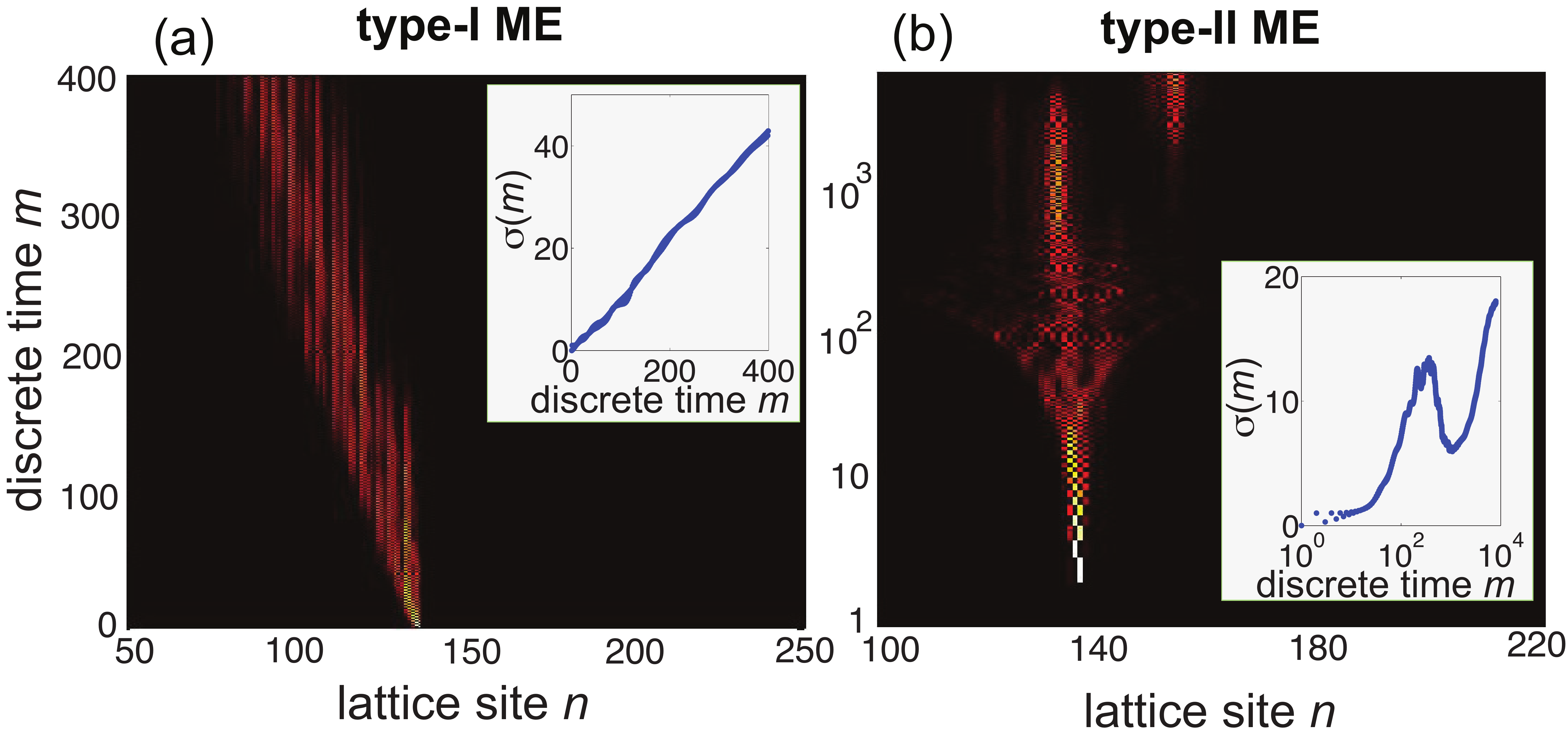}} \caption{ \small
(Color online) Wave spreading in a NH discrete-time photonic quantum walk with (a) type-I, and (b) type-II ME. Parameter values are the same as in Fig.3. The panels show on a pseudo color map the discrete temporal evolution of the normalized occupation probabilities $|\tilde{\psi}_n^{(m)}|^2$ for single-site pulse excitation of the lattice. The insets depict the corresponding behavior of the second-order moment $\sigma(m)$. Note in (b) the use of a log scale for the discrete time $m$.}
\end{figure} 

{\em Conclusions.} Mobility edges in Hermitian crystals with an incommensurate potential enable rather generally transport in the lattice. However, such a scenario is deeply modified when considering NH quasicrystals, where two different types of topological mobility edges can be created by either application of an imaginary gauge field (type-I ME) or by complexification of the phase of the incommensurate potential (type-II ME). Here we unravelled that the different nature  of ME, characterized by different topological numbers, can be detected from the transport features of the system: while in type-I ME ballistic transport is observed, in type-II ME a regime of pseudo dynamical localization is found. We illustrated such a behavior by considering discrete-time photonic quantum walks in NH synthetic mesh lattices with a bichromatic incommensurate potential, which displays either type-I or type-II ME by proper tuning of parameters. The present results provide major insights into the physics of mobility edges in NH quasicrystals, suggesting a dynamical-based approach to distinguish the different topology of NH ME feasible for an experimental demonstration using synthetic photonic mesh lattices.\\

\par

\noindent
{\bf Disclosures}. The author declares no conflicts of interest.\\
\\
{\bf Acknowledgment}. The author acknowledges the Spanish
State Research Agency, through the Severo-Ochoa and Maria de
Maeztu Program for Centers and Units of Excellence in R\&D
(Grant No. MDM-2017-0711).\\
\\
{\bf Data Availability.} No data were generated or analyzed in the presented
research.\\
\\
{\bf Supplemental document}. See Supplement 1 for supporting content.

\end{document}